\documentclass[aps,prl,showpacs,onecolumn]{revtex4-1}
\usepackage[utf8]{inputenc}
\usepackage[english]{babel}
\usepackage{graphicx}
\usepackage{amsmath}
\usepackage{amsfonts}
\usepackage{amssymb}

\begin{document}

\title{From Markovian to non-Markovian persistence exponents}
\date{\today}
\author{Julien Randon-Furling}
\email{Julien.Randon-Furling@univ-paris1.fr}
\affiliation{SAMM (EA 4543), Université Paris-1 Panth\'eon-Sorbonne,\\ Centre Pierre Mend\`es-France, 90 rue de Tolbiac, 75013 Paris, France}

\begin{abstract}

We establish an exact formula relating the survival probability for certain Lévy flights (viz. asymmetric $\alpha$-stable processes where $\alpha = 1/2$) with the survival probability for the order statistics of the running maxima of two independent Brownian particles. This formula allows us to show that the persistence exponent $\delta$ in the latter, non Markovian case is simply related to the persistence exponent $\theta$ in the former, Markovian case via: $\delta=\theta/2$.

Thus, our formula reveals a link between two recently explored families of anomalous exponents: one exhibiting continuous deviations from Sparre-Andersen universality in a Markovian context, and one describing the slow kinetics of the non Markovian process corresponding to the difference between two independent Brownian maxima.
\end{abstract}

\maketitle

\section{Introduction}

In many applications of stochastic modeling, a key question regards the time that a random process will ``survive'' before reaching (or crossing, if the process has jumps) a certain point or absorbing boundary, away from which it started. Examples abound, from spin dynamics~\cite{BBDG} to financial markets \cite{MajBou}, reaction kinetics \cite{ben}, biological systems \cite{goel} or climate science~\cite{BBH}, to name but a few \cite{Redbook}.

Such survival times are usually described by the exponent governing the power-law decay of the long-time asymptotics for the first passage or first crossing time (FPT) density~\cite{BBDG}. They are linked with the time at which the maximum is achieved~\cite{Kn,Yor} and with excursions (fluctuations) away from the average~\cite{BCC}. In a handful of cases, the probability density functions (pdf) associated with the FPT at $0$ can be computed exactly, \textit{e.g.} for Brownian motion starting at $x_0>0$ with diffusion constant $D$, where the method of images yields the well-known density: $f_{D}(x_0,\tau)= x_0 (4\pi D \tau^{3})^{-1/2}\exp (-x_0^2/4D\tau)$, corresponding to a persistence exponent $\theta=1/2$.

As a consequence of E.~Sparre Andersen's celebrated theorem~\cite{SA1,SA2}, this value of $1/2$ expresses a universal behaviour that extends to the more general case of \textit{symmetric} continuous time Markov processes~\cite{ZK,Redbook,Chech1,Koren}, even when the method of images breaks down~\cite{Chech1}. In the case of skewed processes, however, the situation is more complex and offers a larger variety of exponents~\cite{KLCKM}. De Mulatier~\textit{et al}~\cite{deMul} showed recently that the persistence exponent associated with the survival probability $P_{LF}^{(x_0)}$ for asymmetric L\'evy flights with stability index $\alpha \neq 1$ and asymmetry parameter $\beta \in [-1,1]$ can be computed exactly: the result is a family of exponents $\theta(\alpha,\beta)$, which continuously deviates from the Sparre Andersen value of $1/2$ (recovered for $\beta =0$).

\begin{figure}
\includegraphics[scale=0.25]{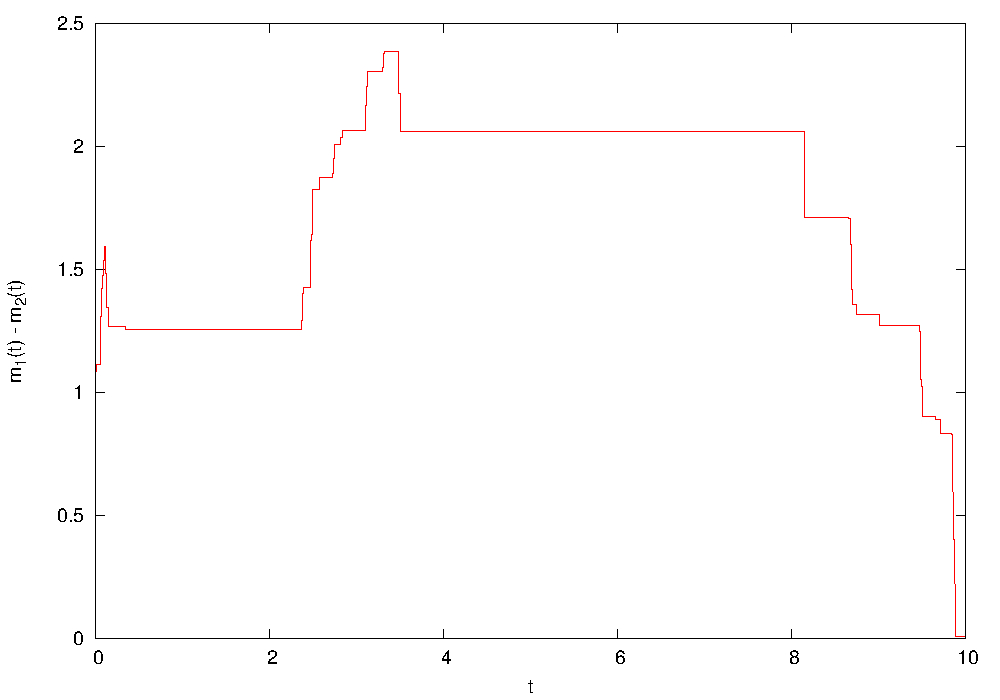}
\caption{\label{CHBMCP}Sample path, until it first crosses $0$, of the non Markov process $m_1(t)-m_2(t)$, where $m_1$ and $m_2$ are the running maxima of two independent Brownian walkers.}
\end{figure}

In the case of non Markov processes, persistence exponents have been studied especially in connection with non-equilibrium dynamics and turbulence~\cite{MMMTA,OCB,MAJCS,PaulS,Henkelbook,PRMP}. Particular interest has been devoted to dynamical phenomena involving extreme-value processes~\cite{MSBC,DHZ,Sire,MRZ,BMaS,MaZi,BNK2}. In a Letter published earlier this year~\citep{BNK}, Ben-Naim and Krapivsky computed the persistence exponent associated with the order statistics of two Brownian maxima (see Fig.~\ref{CHBMCP}). They showed that if $m_1-m_2=a>0$ initially, the probability $P_{BM}^{(a)}(t)$ that the maxima have remained ordered in the same way up to a time $\tau$, decays, when $\tau$ is large, like $\tau^{-\delta}$, where $\delta$ is linked to the diffusion constants of the Brownian particles as follows: $\delta (D_1,D_2)=(1/\pi)\arctan (\sqrt{D_2/D_1})$. In particular, when $D_1=D_2$, $\delta=1/4$.

In this paper, we establish a formula that relates the survival probability of certain L\'evy flights (therefore Markov processes) studied by de Mulatier~\textit{et al.}, with the survival probability examined by Ben-Naim and Krapivsky in a non Markovian context:
\begin{eqnarray}
P_{BM}^{(a)}(t)&=& 2\, \int_{a}^{\infty}dm\ p_{D_1}(a,m;t)\, \times \int_{0}^{\infty}ds\  f_{D_2}(a,s)\, P_{LF}^{(s)} (m). \label{MainFor}
\end{eqnarray}
($f_D$ is the FPT density for Brownian motion, and $p_D$ is the standard Brownian propagator.)

This formula allows us to show how the persistence exponents in both cases are directly related via:
\begin{equation}
\delta (D_1,D_2)=\theta(\alpha,\beta)/2, 
\end{equation}
with $\alpha =1/2$ and $\beta= \left(\sqrt{D_1}-\sqrt{D_2}\right)/\left(\sqrt{D_1}+\sqrt{D_2}\right)$. Thus, our formula also reveals how the $1/4$-exponent obtained in \citep{BNK} when $D_1=D_2$, can be derived directly from the Sparre-Andersen universal behaviour of an underlying, hidden Markov process governing the order statistics of the two Brownian maxima.
\medskip

A key point on which we base our reasoning is the fact, noted by P.~Lévy \cite{Lev}, that the path of the running-maximum process, $M(t)$, of a Brownian motion, is strictly increasing and therefore admits a reciprocal function. This function corresponds to the path of a process, $T(x)$, which is also strictly increasing. But contrary to $M(t)$, $T(x)$ is Markovian. We shall first recall some details about the $T$ process and about general L\'evy-Brownian processes, before establishing our results.

\section{Hitting times of new maxima}
\label{FHT}

Consider a Brownian motion on the real line, with diffusion constant $D$ and with position at time $t$ given by $B(t)$. The transition probability for the motion, sometimes called the free particle propagator, is the well known Gaussian kernel $p_D(x,y;\tau)=(4\pi D \tau)^{-\frac{1}{2}}\exp \left\lbrace -(y-x)^2/4D\tau\right\rbrace$,
which gives the probability density to find the particle in $y$ at time $t+\tau$ given that it was in $x$ at time $t$.

From $B$, the maximum process, that corresponds to the running maximum of the Brownian motion at time $t$, is defined as $M(t)= \mathrm{max} \left\lbrace B(\tau), 0\leq \tau \leq t\right\rbrace$. Using for instance the method of images, one easily shows that the pdf for $M(t)$ is just twice the pdf for $B(t)$, that is $2\, p_D(x_0,M;t)$.

One of the difficulties when working with Brownian maxima is the fact that $M$ is not a Markov process. In some cases, this difficulty can be circumvented by working with $M(t)-B(t)$ which is Markovian (and actually happens to be simply a reflected Brownian motion). In other cases, it can prove useful to rely on a perhaps lesser-known process, associated with the times when new maxima are achieved: $T(x)=\inf \left\lbrace t\geq 0, B(t) = x \right\rbrace$.

Lévy showed that $T$ is Markovian and $\alpha$-stable with $\alpha = 1/2$, since the characteristic function (which is just the Fourier transform of the law) of the random variable $T(x)$ has the form \cite{Lev}:
\begin{equation}
\mathbb{E}\left[e^{iuT(x)}\right]=e^{-(1-i\ \mathrm{sign}(u))x\sqrt{\frac{\mid u \mid}{2D}}}, \label{PhiT}
\end{equation}
where $\mathbb{E}\left[\dots\right]$ denotes probabilistic expectation, and $\mathrm{sign}(u)$ is the sign of $u$. When the diffusion constant $D$ is $1/2$, this process is called the standard stable subordinator of index $1/2$. Its propagator can be expressed easily \cite{MorPer} from the first-passage time density of Brownian motion, and one finds a transition kernel from $s$ to $t>s$ in a ``time'' $a$ given by:
\begin{equation}
f_D(a,t-s)=\frac{a}{\sqrt{4\pi D (t-s)^{3}}}e^{-\frac{a^2}{4D(t-s)}}. \label{PropSub}
\end{equation}
(Recall that here $a$ is the time parameter for the $T$ process, while $s$ and $t$ correspond to positions. Also, note that this is a densitiy in $t$ and therefore will be multiplied by $dt$ upon integration so that, dimension-wise, there is nothing wrong with the denominator.)

Stable subordinators are an extreme case of $\alpha$-stable processes. These processes play an important role in line with the generalized central limit theorem~\cite{Feller}: so-called L\'evy flights, also known as L\'evy-Brownian motions~\cite{DGH}, have emerged as powerful tools to model non Gaussian phenomena~\cite{POC,ZX,PE,JS,RSIB}. They are continuous time Markov processes that can be described in Langevin formalism as generalized Brownian motion, where the driving term, instead of a simple Gaussian white noise is a general Lévy noise $\zeta$, that induces independent increments identically distributed according to an $\alpha$-stable law~\cite{DGH}: $\dot{x}(t)=\zeta (t)$. Such stable laws are specified by four parameters \cite{Apple}: their stability index $\alpha \in]0,2]$, a scale parameter $\sigma$, a location parameter $\mu$, and a skewness parameter~$\beta$. For $\alpha \neq 1$, the characteristic function of their position at time $t$ admits the following L\'evy-Khintchine general form~\cite{Koren}:
\begin{equation}
\phi _t (u)=e^{i\mu u -\sigma ^{\alpha} t \mid u \mid ^{\alpha} \left(1- i \beta \tan \left(\frac{\pi}{2}\alpha \right) \mathrm{sign}(u)\right)}.\label{LK}
\end{equation}
Standard Brownian motion corresponds to $\alpha =2$, the Cauchy-Lorentz process is $\alpha=1, \beta=0$, and the L\'evy-Smirnov driven process, with $\alpha =0.5, \beta=1$, is the same as the stable subordinator mentioned above.

\section{Time-lag process}
\label{TLP}

Let now $B_1$ and $B_2$ be two independent Brownian walkers. Write $m_1$ and $m_2$ for their running maxima and write $T_1$ and $T_2$ for the associated hitting-time processes as defined in the previous section. We set $Z(x) = T_2(x)-T_1(x)$ and call $Z$ the \textit{time-lag process}: indeed, if $Z(x)=\tau$, it means that, compared to $B_1$, $B_2$ first hit $x$ with a delay $\tau$.

The important point here is that if $m_1$ and $m_2$ switch order, so will $T_1$ and $T_2$, and therefore $Z$ will change signs. One can therefore compute the probability that $m_1$ and $m_2$ remain in the same order up to time $t$ by examining the probability that $Z$ remains positive on the interval $\left[0,m_1(t)\right]$ (of course, one will need to pay attention to the fact that the right-hand side of this interval is itself a random variable).

Let us first determine the characteristic function of $Z(x)$, in terms of the diffusion constants $D_1$ and $D_2$. Using the independence of $T_1$ and $T_2$, one has:
\begin{equation}
\mathbb{E}\left[e^{iuZ(x)}\right]=\mathbb{E}\left[e^{iuT_2(x)}\right]\mathbb{E}\left[e^{-iuT_1(x)}\right].
\end{equation}
Substituting formula~(\ref{PhiT}) for the characteristic functions on the right-hand side yields:
\begin{eqnarray}
\mathbb{E}\left[e^{iuZ(x)}\right]&=&\exp \left\lbrace-\left(1- i \frac{\sqrt{D_1}-\sqrt{D_2}}{\sqrt{D_1}+\sqrt{D_2}}\ \mathrm{sign}(u)\right)\ \left(\frac{\sqrt{D_1}+\sqrt{D_2}}{\sqrt{2 D_1 D_2}}\right) x\sqrt{\mid u \mid}\right\rbrace. \label{PhiZ}
\end{eqnarray}

From this characteristic function and~(\ref{LK}), one can readily identify an $\alpha$-stable process with $\alpha=1/2$, and asymmetry coefficient $\beta= \left(\sqrt{D_1}-\sqrt{D_2}\right)/\left(\sqrt{D_1}+\sqrt{D_2}\right)$. We immediately notice that when the diffusion constants are the same, $\beta$ will be zero and therefore $Z$ will simply be a symmetric L\'evy flight. We will make use of this observation in section~5, but let us first derive a general formula.

\section{General formula}
\label{GF}

Let $B_1$ and $B_2$ have diffusion constants $D_1$ and $D_2$, and let $B_1(0)=a$ with $a>0$, and $B_2(0)=0$. In this case, $T_1(x)$ is not defined for $x<a$, but $T_1(a)=0$. Also $T_2(0)=0$ and $T_2(a)$ is distributed according to the transition kernel (\ref{PropSub}) ---~so it has almost surely some strictly positive value. Hence, $Z(x)=T_2(a+x)-T_1(a+x)$ is a $1/2$-stable L\'evy flight that starts at some value $T_2(a)>0$.

Now we wish to encode the fact that the time-lag process does not have any zero crossing before ``time'' $m_1(t)$ (recall that the time space for the $Z$ process is the position space for the Brownian particles). This is saying that the first time at which $Z$ crosses $0$ should be larger than $m_1(t)$ and it can be expressed in terms of the survival probability $P_{LF}\left(m_1(t)\right)$, provided that the initial value, $T_2(a)$, is averaged out using $f_{D_2}(a,s)$, the propagator from (\ref{PropSub}):
\begin{equation}
R(m_1,a)=\int_{0}^{\infty}ds\  f_{D_2}(a,s)\, P_{LF}^{(s)}(m_1(t)).\label{fm1a}
\end{equation}

We can now write our general formula for $P_{BM}(t)$ by integrating~(\ref{fm1a}) over $m_1(t)$:
\begin{equation}
P_{BM}^{(a)}(t)= \int_{a}^{\infty}dm\ 2\, p_{D_1}(a,m;t)\, R(m,a).\label{PSurv}
\end{equation}
Of course, $P_{BM}(t)$ depends on $a$, the initial distance between the two walkers. Also, to compute $P_{BM}(t)$ more explicitly, one would need to know the exact form of $\rho_{s}(x)$, the pdf for the time at which a $1/2$-stable L\'evy flight with characteristic function as in~(\ref{PhiZ}) first crosses $0$, having started at $s$. This would allow the computation of $P_{LF}^{(s)}$. Unfortunately, such pdfs are known explicitly only for very few Lévy processes: as soon as jumps are not negligible (as is the case for  $1/2$-stable processes), hitting times and crossing times differ, and standard methods like the method of images fail~\cite{Chech1,Koren}. However, one can extract from~(\ref{PSurv}) a non-trivial relationship between some Markovian persistence exponents and some non Markovian ones, as we shall now see.

\section{Bijection between exponents}

In the general case, when the diffusion constants are different, the time-lag process $Z$ will be $1/2$-stable but asymmetric, with a skewness coefficient $\beta= \left(\sqrt{D_1}-\sqrt{D_2}\right)/\left(\sqrt{D_1}+\sqrt{D_2}\right)$. Such asymmetric L\'evy flights are found to have power-law tailed persistence~\cite{deMul}, so that if one writes $\theta(1/2,\beta)$ (or simply $\theta $ here) for their persistence exponents, one can express in terms of these the exponent governing the asymptotics of $P_{BM}(t)$. Indeed, from~(\ref{fm1a}) and (\ref{PSurv}):
\begin{eqnarray}
&&  P_{BM}(t)= \int_{a}^{\infty}dm\  2\, p_{D_1}(a,m;t)\ \int_{0}^{\infty}ds\   f_{D_2}(a,s)\,  \int_{m}^{\infty}dx\ \rho_{s} (x)\nonumber \\
  &=&\int_{0}^{\infty}ds\  f_{D_2}(a,s)\, \int_0^\infty dx\  \rho_{s} (x+a)\ \mathrm{erf}\left(\frac{x}{\sqrt{2t}} \right)\nonumber\\
  &=& \sqrt{2t} \int_{0}^{\infty}ds\  f_{D_2}(a,s)\  \int_0^\infty dx\ \rho_s (x\sqrt{2t}+a)\  \mathrm{erf}\left(x\right).\nonumber
   \end{eqnarray} 
And, using $\rho_{s} (v) \sim h(s)\, v^{-(1+\theta)} $ when $v$ is large, with $h$ a function of the initial point $s$, one finds that:   
\begin{eqnarray}
&&  P_{BM}(t) \sim \sqrt{2t}  \int_{0}^{\infty}ds\  f_{D_2}(a,s)\ \int_0^\infty dx\  h(s)\ (x\sqrt{2t}+a)^{-(1+\theta)}\  \mathrm{erf}\left(x\right) \nonumber \\
   &\sim & (\sqrt{2t})^{-\theta} \int_{0}^{\infty}ds\  h(s)\ f_{D_2}(a,s)\, \int_0^\infty dx\  x^{-(1+\theta)}\  \mathrm{erf}\left(x\right)\nonumber \\
    &\sim &\ \tilde{h}(a)\ t^{-\frac{\theta}{2}}. \label{ExpSBM}
    \end{eqnarray}  
(We have used the fact that any finite $a>0$ will become negligible compared to $\sqrt{2t}$ when $t$ becomes large enough, and we write $\tilde{h}$ for the function obtained from the remaining integrals.)

Thus, we obtain our second main result:
\begin{equation}
P_{BM}(t)\sim t^{-{\theta(\beta)/2}}. \label{ExpTheta}
\end{equation}

To confirm this result, note that we know from~\cite{deMul} the exact value of $\theta(\beta)$: this is given by $1/2 - 2/\pi \arctan (\beta)$. It is then easy to see, by means of a simple trigonometric identity, that with $\beta$ as stated in~(\ref{PhiZ}) in terms of $D_1$ and $D_2$, one finds: $\theta(\beta)/2=(1/\pi) \arctan \left(\sqrt{D_2/D_1}\right)$, in full agreement with~\cite{BNK}.

Also, in simple, extreme cases, one recovers intuitive results: \textit{eg} for $D_1 \gg D_2$, one expects $P(t)$ to be equal to $1$ as the running maximum of the first particle will keep being too high for the second particle to exceed it. In this case, $\beta\rightarrow 1$ and one sees from formula~(\ref{PhiZ}) that the time-lag process reduces to a $1/2$-stable subordinator, that is a jump process with purely positive jumps, so that $Z$ never crosses $0$ and $P(t)=1$ indeed.

\section{Hidden Sparre Andersen behaviour}
\label{HSA}

An interesting phenomenon is to be noticed in the special case when the two Brownian motions have the same diffusion constant. Then, $\beta =0$ and the time-lag process becomes \textit{symmetric}. Therefore, the pdf for its first crossing time will obey the universal Sparre Andersen behaviour so that  $\rho_{s}(x) \sim h(s) x^{-3/2}$, and equivalently $\theta(\beta)=1/2$. Our second main result, Equation~(\ref{ExpTheta}), therefore sheds light on how the anomalous $1/4$-exponent computed in~\cite{BNK} emerges from the underlying ``standard'' Sparre Andersen behaviour of the time-lag process.

\section{Conclusion}
We have established an exact formula for the survival probability $P_{BM}(t)$ of the order of two Brownian maxima, by linking it with the survival probability of an underlying L\'evy flight. This yields a bijection between non trivial persistence exponents. Via the results in \citep{BNK}, our formula also establishes a link between persistence exponents for asymmetric $1/2$-stable Lévy flights and an eigenvalue of the angular component of the 2-dimensional Laplace operator.

One of the main features of the method used in this paper is the fact that it maps a non Markov problem onto a Markov one, moving from Brownian maxima to Lévy flights. The extent to which the same mapping may be used to obtain other results, in this and other contexts (order statistics of more than two maxima, higher dimension,\dots), should and will be investigated.

\bibliography{BMax}

\end{document}